\documentclass[
                breakaddress,  
]{nrc2}                          

    \usepackage{graphicx}       

    \usepackage{amsmath}        
    \usepackage{amssymb}        
    \usepackage{bm}             

\volyear{XX}{2014}               
\journal{Can. J. Phys.}                       
\journalcode{cjp}                   
\filenumber{xx}                    

\received{}                      
\revreceived{}                   
\accepted{}                      
\revaccepted{}                   

\begin{document}

\title{The Quantum Echo of the  Early Universe}

\shortauthor{ Blasco, Garay, Mart\'in-Benito, and Mart\'in-Mart\'inez} 

\author{Ana Blasco}
\email{anablasc@ucm.es}
\address{Departamento de F\'isica Te\'orica II, Universidad Complutense de Madrid, 28040 Madrid, Spain.}

\author{Luis J. Garay}
\address{Departamento de F\'isica Te\'orica II, Universidad Complutense de Madrid, 28040 Madrid, Spain.}

\author{Mercedes Mart\'in-Benito}
\email{M.Martin@hef.ru.nl}
\address{Radboud University Nijmegen,
Institute for Mathematics, Astrophysics and Particle Physics, NL-6525 AJ Nijmegen, The Netherlands.}

\author{Eduardo Mart\'in-Mart\'inez}
\email{emartinm@uwaterloo.ca}
\address{Institute for Quantum Computing, University of Waterloo, Waterloo, Ontario, N2L 3G1, Canada.\\
Department of Applied Mathematics, University of Waterloo, Waterloo, Ontario, N2L 3G1, Canada.\\
Perimeter Institute for Theoretical Physics, Waterloo, Ontario N2L 2Y5, Canada.}

\begin{abstract}
We show that the fluctuations of quantum fields as seen by late comoving observers are significantly influenced by the history of the early Universe, and therefore they transmit information about the nature of spacetime in timescales when quantum gravitational effects were non-negligible. We discuss how this may be observable even nowadays, and thus used to build falsifiability tests of quantum gravity theories.
\end{abstract}

 \maketitle			

After more than 70 years of laborious attempts, a keystone of the physical understanding of the Universe still remains elusive: The description of the microscopic structure of space-time, or in other words, the formulation of a complete theory of quantum gravity. 
In order to test whether a particular quantum gravity theory may be correct,  we need  to make falsifiable predictions, and thus we need  experimental evidence of quantum gravity effects so we can compare theory and observations. This poses a serious challenge as to detect these effects we might need to go all the way down to the Planck scale, which is out of reach of any earthbound experiment. Fortunately, there is a window to quantum gravitational  effects that is becoming more and more accessible. This window is cosmology,  and in particular the possible evidence of early Universe physics  that might have been imprinted onto the cosmic microwave background (CMB). 

We live exciting times for the field of cosmology. Very recently the BICEP2 experiment has reported results on how  B-modes, which may have originated on primordial gravitational waves (See, e.g. \cite{TanViejoQueNoHabiaNiGimnasio,AntesDeEntrenarBebeMuchaAgua,AntesDeSacarBicepEntrena}), affected the polarization of the CMB \cite{SacaBicepf,SacaBicep2}.
Although doubts have been cast about the primordial origin of the B-modes reported to be found in \cite{SacaBicepf}, primordial gravitational waves are fluctuations of the geometry, so these experiments may pave the way to exploring quantum gravity effects in the near future.

In this context a natural question arises: have  quantum gravity signatures really survived from the early Universe until the current era? If so, how strong are they? Will we be able to validate (or falsify) different quantum gravity proposals by looking at the data? In this essay we explore a simple way, based on a toy model, to assess the strength of the quantum signatures of  the  early Universe that might be observed nowadays.

Our approach consists of determining the response of an idealized particle detector which has remained coupled to matter fields from the early stages of the Universe until today. We pose the question whether this detector would conserve any information from the time when it witnessed the very early Universe dynamics. Let us reflect for a second upon the comparison of these two outrageously different timescales:  Plank time and the age of the Universe. Intuitively, one would think that  any effect imprinted on the response of the detector in the early Universe would most likely have been already washed out, and hence there is little hope in finding any trace of early Universe physics in the response of the particle detector today. Surprisingly  this intuition is wrong.

To show this, we compare  the response of the detector evolving under two different Universe dynamics which disagree only during the short time when matter-energy densities are of the order of the Planck scale. As a convenient first example, we choose to compare classical General Relativity (GR) with effective Loop Quantum Cosmology (LQC) \cite{Bojowald:2008zzb,Ashtekar:2011ni,Banerjee:2011qu}.  We consider a spatially flat, homogeneous and isotropic Universe,\\  
$ ds^2=~a(t)[-d\eta(t)^2+d \bm{x}^2]$, with  a massless scalar $\varphi$ as matter source. The scale factor for each dynamics reads \cite{ours}
 \begin{align}\label{class}
a_{GR}(t)&=\frac{( \ell_\textsc{p}^2\pi_\varphi^2)^{1/6}}{L}t^{1/3},\\
a_{LQC}(t)&=\frac{l}{L}\left(\frac{\pi_\varphi^2}{{\ell_\textsc{p}^2}}\right)^{1/6}\left[1+\left(\frac{\ell_\textsc{p}^2}{l^3}\right)^2t^2\right]^{1/6},
\end{align}
and the conformal time $\eta$ in terms of the comoving time $t$ is
\begin{align}\label{etaclass}
\eta_\text{GR}(t)&=\frac{3L\,t^{2/3}}{2(\ell_\textsc{p}^2\pi_\varphi^2)^{1/6}},\\
\eta_{LQC}(t)&=\frac{L}{l}\left(\frac{\ell_\textsc{p}^2}{\pi_\varphi^2}\right)^{1/6}t \cdot {_2}F_1\left[\frac1{6},\frac1{2},\frac{3}{2},-\left(\frac{\ell_\textsc{p}^2}{l^3}t\right)^2\right].
\end{align}

\begin{figure}
\begin{center}

\topcaption{Scale factor as a function of the proper time for $l=1$,  $\pi_\varphi=1000$. The dashed blue curve represents the classical scale factor 
$a_\text{GR}(t)L$ and the solid red curve corresponds to the LQC effective scale factor 
$a_\text{LQC}(t)L$. All quantities are expressed in Planck units, i.e.  $\ell_\textsc{p}=1$.}
\includegraphics[width=0.45\textwidth]{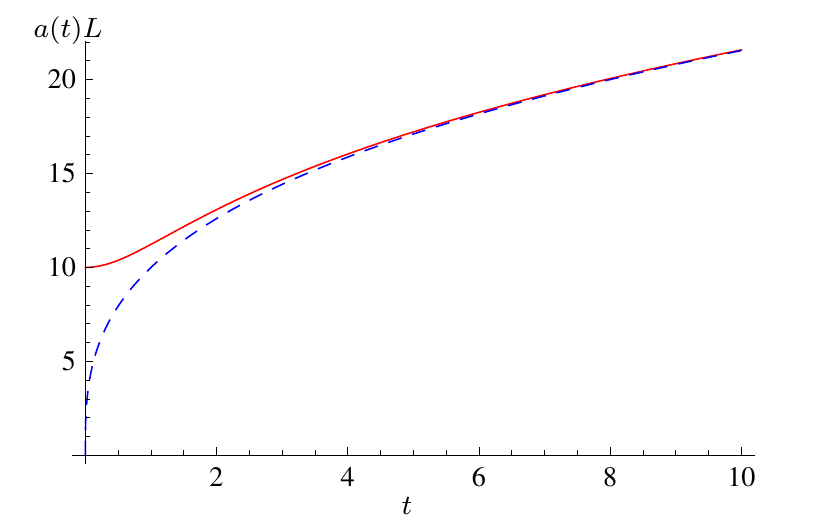}
\label{fig:scalefactors}
\end{center}
\end{figure}
Here, $\ell_\textsc{p}=\sqrt{12\pi G}$ is the Planck length,
$\pi_\varphi$ is the momentum conjugate to $\varphi$ and is a constant of motion, $l$ is a quantization parameter (in LQC the volume has a discrete spectrum  equally spaced by $2l^3$ units)
\cite{Ashtekar:2011ni,Banerjee:2011qu}, and ${}_2F_1$ is an ordinary hypergeometric function.  For simplicity we consider a three-torus spatial topology, with coordinates in the interval $[0,L]$. For compactification scales larger than the observable Universe, this flat topology is compatible with  observations \cite{Mukhanov:2005sc}. The classical GR solution,  $a_{GR}(t)=~\lim_{l\rightarrow 0} a_\text{LQC}(t)$, displays a big-bang singularity at 
 $t=0$. In contrast, $a_\text{LQC}(t)$ never vanishes (see Fig.~\ref{fig:scalefactors}). LQC   replaces this singularity  by a big-bounce, i.e.,  the Universe shrinks for $t<0$, bounces at $t=0$, and expands for $t>0$. In the limit $t\gg{l^3}/\ell_\textsc{p}^2$, $\eta_\text{LQC}(t)\rightarrow \eta_\text{GR}(t)+\beta$ with $\beta=\frac{l^2L\sqrt{\pi}\,\Gamma\left(-\frac1{3}\right)}{(\pi_\varphi^2 
\ell_\textsc{p}^{10})^{1/6}2\Gamma\left(\frac1{6}\right)}$.
We consider a conformal massless scalar field $\phi$  filling the Universe. We quantize this field choosing the conformal vacuum, which remains invariant as the Universe expands \cite{cosmoq,cosmoq2}. 
The proper time of comoving observers (who see an isotropic expansion)  does not coincide with the conformal time. Hence,  comoving detectors actually detect particles even in the conformal vacuum \cite{cosmoq}.  This is the well-known Gibbons-Hawking effect \cite{GibHawking}, which we analyze to identify signatures of quantum gravitational effects on the particle detector.
The Unruh-DeWitt model  \cite{DeWitt} describes the 
local monopole interaction of a two-level quantum system with a scalar field \cite{Birrell,cosmoq,Reznik2005,AasenPRL}. Although simple, this model encompasses the fundamental features of the radiation-matter interaction  \cite{Wavepackets,Alvaro}. We consider an Unruh-DeWitt detector stationary in the comoving frame, $\bm{x}(t)=\bm x_0$, and initially in its ground state. The Hamiltonian of the coupled system  in the interaction picture is $\hat{H}_I(t)=~\lambda\;\chi(t)(\sigma^+ e^{i\Omega t}+\sigma^- e^{-i\Omega t})\hat{\phi}[\bm{x}(t),\eta(t)]$, where $\lambda$ is the coupling strength, $\chi(t)$ is the detector's switching function, which we choose to be analytic. $[\bm{x}(t),\eta(t)]$ represents the detector's world-line, $\Omega$ is its energy gap, and $\sigma^\pm$ are $SU(2)$ ladder operators. Assuming a small enough $\lambda$, we can  compute  the transition probabilities for the detector switched on at $T_0$ to be excited at time $T$ within perturbation theory:
\begin{align} 
P_\text{e}(T_0,T)& =\lambda^2 {\sum_{\bm{n}}}|I_{\bm{n}}(T_0,T)|^2 +\mathcal{O}(\lambda^4),\\ \label{integ}
I_{\bm{n}}(T_0,T)&=\int_{T_0}^T \!\!\! \text{d}t 
\frac{\chi(t) }{a(t)\sqrt{2\omega_{\bm n}L^3}}e^{-\frac{ 2\pi\mathrm{i} 
\bm n\cdot \bm x_0}{L} }e^{\mathrm{i}[\Omega t + \omega_{\bm{n}} 
\eta(t)]}.
\end{align}
 Here, $\omega_{\bm{n}}=\frac{2\pi}{L}\left|{\bm{n}}\right|$, and \mbox{$\bm{n}=(n_x,n_y,n_z)\in\mathbb{Z}^3$.}  We now compare the probability  of the detector to get excited 
when the Universe evolves under both dynamics,  $P^{\text{LQC}}_\text{e}(T_0,T)$, and $P^{\text{GR}}_\text{e}(T_0,T)$:
\begin{align}
&\Delta P_\text{e}(T_0,T)\equiv P^{\text{LQC}}_\text{e}(T_0,T)-P^{\text{GR}}_\text{e}(T_0,T)\\\nonumber
&=\lambda^2{\sum_{\bm n}}\bigg[\left|I_{\bm 
n}^\text{LQC}(T_0,T_\text{m})\right|^2-\left|I_{\bm n}^\text{GR}(T_0,T_\text{m})\right|^2 
\\\nonumber
& +2 \text{Re}\Big({I_{\bm n}^\text{GR}}^*(T_\text{m},T)\Big[e^{-\mathrm{i}\beta\omega_{\bm n}}I_{\bm 
n}^\text{LQC}(T_0,T_\text{m})\\\nonumber
&-I_{\bm n}^\text{GR}(T_0,T_\text{m})\Big]\Big)\bigg]+\mathcal{O}(\lambda^4).
\end{align}
$T_\text{m}\in (T_0,T)$ is a short time sufficiently large for
$\eta_\text{LQC}(T_\text{m})\approx\eta_\text{GR}(T_\text{m})+\beta$, 
typically  few times $l^3/\ell_\textsc{p}^2$, used to split the integrals \eqref{integ} into two intervals:  a tiny interval $t\in [T_0,T_\text{m}]$ (comparable to the Planck scale) where LQC and GR  appreciably predict different dynamics;   and $t\in [T_\text{m},T]$ where both dynamics are essentially the same.  
The first shocking observation is that the difference of the detector's particle counting in both scenarios, 
$\Delta P_\text{e} (T_0,T)$, will be considerable, even for 
$T\gg l^3/\ell_\textsc{p}^2$, that is, if we look at the detector 
nowadays.

Let us study how sensitive  the response of the detector is to the LQC quantum parameter $l$  
that characterizes the quantum of volume. With this aim we use as estimator  the mean relative difference between probabilities of excitation averaged over a long interval in the late time regime  $\Delta T=T-T_{\text{late}}$, with $\Delta T,\;T_\text{late}\gg l^3/\ell_\textsc{p}^2$:
\begin{align}\label{estimator} 
E=\left\langle\frac{\left\langle  \Delta P_\text{e}(T_0,T) 
\right\rangle_{\mathcal{T}}}{\left\langle  P^\text{GR}_\text{e}(T_0,T)
\right\rangle_{\mathcal{T}}}\right\rangle_{\Delta T}.\end{align}
This estimator tells us the difference in 
magnitude between the number of clicks of the detector in the GR and LQC backgrounds. The internal averages  in \eqref{estimator}  are given by
\begin{equation}\label{ave}
\left\langle   P_\text{e}(T_0,T) \right\rangle_{\mathcal{T}}=\frac{1}{\mathcal{T} }\int^T_{T-\mathcal{T}} P_\text{e}(T_0,T')\,\text{d}T',
\end{equation}
where $\mathcal{T}\gg l^3/\ell_\textsc{p}^2$ is the time resolution with which we can probe the 
detector.  This is so because it is not possible to resolve times as small as $l^3/\ell_\textsc{p}^2$ 
(roughly, the  Planck scale), so any observation  made on particle 
detectors today will necessarily be averaged over many Planck times.  Moreover, in order to remove any 
possible spurious effects of the big differences in the  
scales of the problem, we will consider a particle detector with a subplanckian energy gap 
$\Omega\ll\ell_\textsc{p}^2/l^3$.   One may legitimately wonder if these practical considerations may destroy the early Universe signal. Indeed, these constraints 
 partially erase the observable difference between the response of the 
detector in the two scenarios.

\begin{figure}
\begin{center}
\topcaption{Logarithmic plot of the relative difference of the averaged probabilities, E, as a function of the parameter $l$, for $\Omega\ll \ell_\textsc{p}^2/l^3$ and $\pi_\varphi=1000$.  The detector is switched on at $T_0=0.01$ (some early time after the bounce). The variation of the response of the detector (the intensity of Gibbons-Hawking type quantum fluctuations) grows exponentially with l.}
\includegraphics[width=0.45\textwidth]{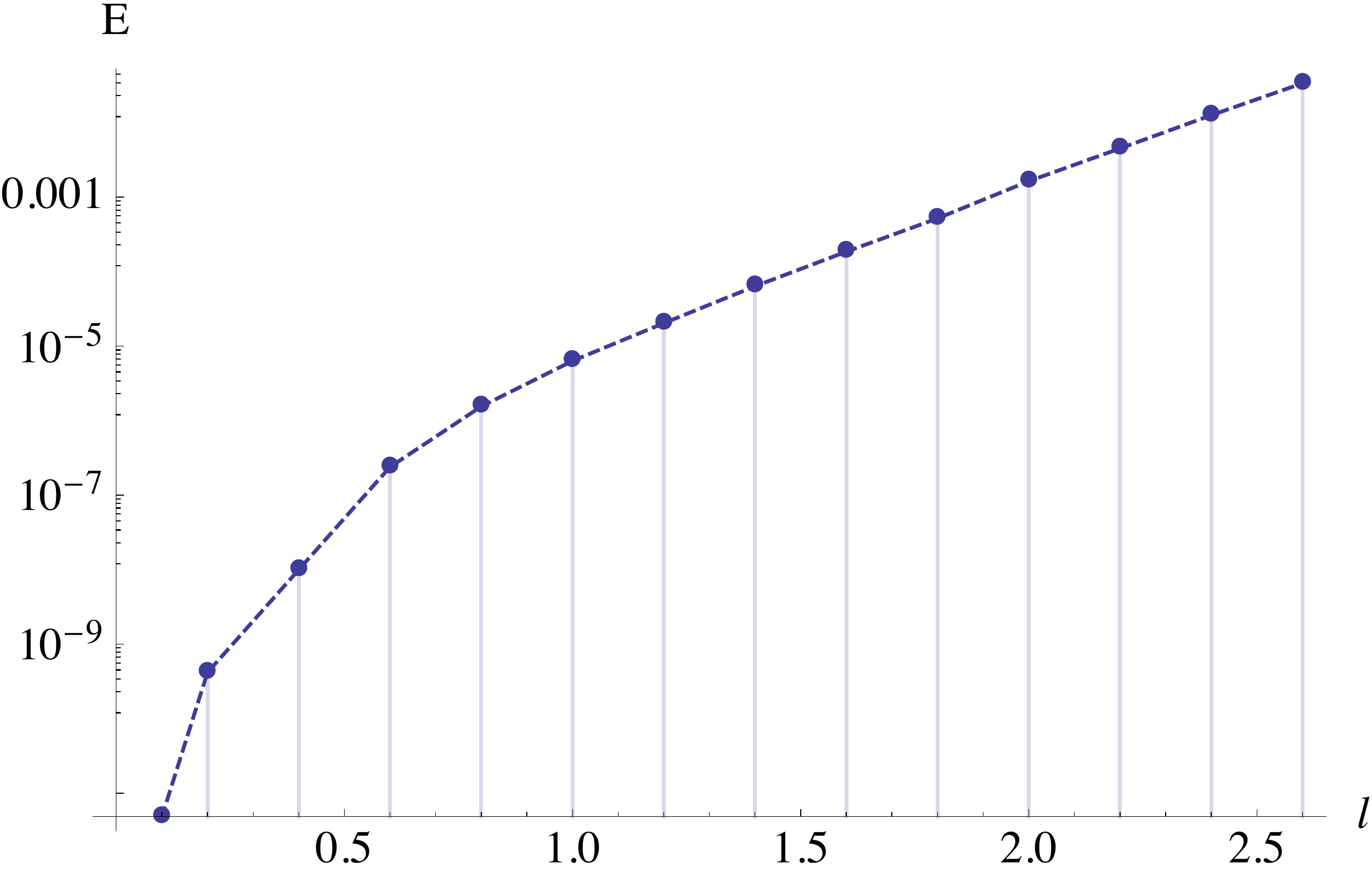}
\label{fig:tomajeroma}
\end{center}
\end{figure}

 Remarkably, the difference between the long time averaged response of highly
sub-Planckian detectors in the GR and LQC scenarios \eqref{estimator} remains non-negligible even 
under these coarse-graining conditions.  As shown in Fig. \ref{fig:tomajeroma}, the variation of the response of the detector (the 
intensity of Gibbons-Hawking-type quantum fluctuations) grows exponentially with the size of the quantum of length 
$l$. This exponential trend does not depend on the timescale of the detector's activation or the nature of  the switching function \cite{ours}. In consequence, $l$ cannot be much beyond the Planck scale or the effects would be too large nowadays. 
This suggests that cosmological observations could put stringent upper bounds to  the size of the quantum of volume in LQC or, equivalently, to the time scale $T_\text{m}$ when the 
quantum dynamics corrections become  negligible.

Although this is a toy model, it captures the essence of a key 
phenomenon:  Quantum field fluctuations are extremely sensitive to the physics of 
the early Universe. More importantly, the signatures of these fluctuations survive in the current era. 

We would like to emphasize that  the use of LQC in this derivation is anecdotical, we assert that our result is far more general: We use LQC as a convenient example of early Universe dynamics different from GR.  One would expect that in the early Universe, even if a quantum gravity theory predicts that there is no such thing as  spacetime geometry near the Plank scale, there will  be intermediate energy scales where a semi-classical theory is applicable, producing effective quantum corrections to the Friedmann equations. This will predict, for a short time, a spacetime dynamics which deviates from GR through quantum corrections. Our main result prevails: The response of a particle detector today carries the imprint of the specific dynamics of the spacetime in the early Universe.

Current research  derived from these results aims towards extending this methodology to further explore the window opened by  the combination of cosmology and quantum information, to study the possibility of optimal transmission and recovery of  information propagated through cosmological catastrophes, such as the big-bang, inflation or a quantum bounce.

\section{Acknowledgements}
L.J.G. and M.M-B. acknowledge financial support from the MICINN/MINECO Projects No. FIS2011- 30145-C03-02 and FIS2014-54800-C2-2-P. 
M.M-B acknowledges financial support from the Netherlands Organisation for Scientific Research (NWO) (Project No. 62001772).

\bibliographystyle{unsrt}
\bibliography{references}

\end{document}